\documentclass[pra, reprint, showkeys, amsmath,amssymb, aps,]{revtex4-2}
\usepackage[compatibility=false]{caption}
\usepackage{caption}
\usepackage{graphicx}
\graphicspath{{./media/}}
\usepackage{subcaption}

\usepackage[colorlinks=true, linkcolor=blue, citecolor=blue, urlcolor=blue]{hyperref} 
\usepackage{placeins}

\begin{document}
\title{Shear-induced pressure anisotropy in granular materials of nonspherical particles}

\author{Huzaif Rahim}
\email{huzaif.rahim@fau.de}
\author{Sudeshna Roy}
\author{Thorsten P\"oschel}

\affiliation{Institute for Multiscale Simulation, \\ Friedrich-Alexander-Universit\"at Erlangen-N\"urnberg, \\ Cauerstrasse 3, 91058 Erlangen, Germany}

\begin{abstract} 
When a granular material composed of elongated grains is sheared in a split-bottom shear cell, a pressure difference develops within the material. This pressure difference depends on the interparticle friction ($\mu$), which affects shear localization and particle alignment. For large $\mu$, alignment is confined to a narrow shear band, leading to localized increases in packing density and pressure. For small $\mu$, particles align over a wider region, leading to a nearly uniform packing density and pressure throughout the material. In contrast, spherical particles, regardless of $\mu$, maintain a uniform packing density and pressure throughout the material. We observe a phenomenological similarity to the Weissenberg effect in non-Newtonian fluids, where normal stress differences induce radial pressure gradients, unlike the uniform pressure in Newtonian fluids.
\end{abstract}

\maketitle 

\section{Introduction}
Granular materials generally exert non-hydrostatic pressures, in contrast to fluids, where the isotropic pressure increases linearly with depth \cite{jaeger1996granular}. In confined geometries such as silos or hoppers, wall friction supports part of the vertical load, leading to pressure saturation known as the Janssen effect \cite{aguilar2025janssen, thorens2021taming}. In granular materials, the pressure is determined by the contact forces between particles \cite{weinhart2016influence, singh2015role}. For nonspherical particles such as elongated grains, these contact forces --and therefore the mechanical response-- are governed by microstructural rearrangements that depend on particle aspect ratio ($\text{AR}$), orientation, and interparticle friction \cite{anthony2005influence, yu1998prediction, cheng2000dynamic}.

Shearing, shaking, or pouring such elongated grains induces orientational ordering, which affects packing density, surface profile, and stress distribution \cite{nagy2023flow, pol2022kinematics, campbell2011elastic, rahim2024alignment}. In a split-bottom shear cell, these elongated grains align along the shear direction \cite{rahim2024alignment, borzsonyi2012orientational}. This alignment competes with Reynolds dilatancy \cite{wegner2014effects, hosseinpoor2021rheo}, which describes the shear-induced volume expansion of a granular packing \cite{reynolds1885lvii}. Dilatancy leads to heap formation and secondary flows \cite{krishnaraj2016dilation, wortel2015heaping, dsouza2017secondary, fischer2016heaping}, while alignment causes depression above the shear band \cite{wegner2014effects, rahim2024alignment}. 
The resulting surface morphology depends on particle shape, friction, and initial packing \cite{rahim2025impact, rahim2024alignment}.

The shear band is the focal region for microstructural evolution under shear, where the alignment of elongated particles increases contact density and induces stress anisotropy \cite{fu2011fabric, liu2018application, nagy2023flow, pol2022kinematics, rahim2024alignment}. This anisotropy leads to spatial variations in pressure \cite{liu2018application}. The effect is stronger for particles with larger aspect ratio, which form highly ordered, anisotropic contact networks \cite{rahim2024alignment, liu2024effect, hidalgo2009role, kanzaki2011stress}. Spherical particles, on the other hand, do not align and therefore maintain nearly isotropic stresses under shear \cite{rahim2024alignment, zhu2023discrete}. Although shear-induced dilation and compaction have been widely studied \cite{sperl2006experiments, lei2019side, vescovi2016merging}, their influence on the pressure field in sheared nonspherical granular materials remains unexplored.

The linear split-bottom shear cell (LSC) is used to study the effect of particle shape and material properties on granular behavior, including secondary flows, orientational order, and rheological properties such as stress distribution, effective friction, and apparent viscosity \cite{dijksman2010granular, fenistein2003wide, fischer2016heaping, wegner2014effects}. Despite its limited size, the periodic boundary along the flow direction mimics an infinitely long system \cite{ries2007shear, dsouza2021dilatancy,roy2021drift, rahim2024alignment}.

We use Discrete Element Method (DEM) simulations in the LSC geometry to investigate how particle shape, characterized by the aspect ratio and interparticle friction ($\mu$) affect the pressure distribution in sheared granular material. The simulation results for systems composed of spherical ($\text{AR}=1$) and elongated particles ($\text{AR}=5$), with $\mu \in [0.01, 0.8]$ are analyzed and compared. By examining particle alignment and the resulting stress fields, we show how particle shape and friction determine the pressure variation within and around the shear band.

\section{Numerical setup and methods}

\subsection{System geometry}
\begin{figure}[!htbp]
\centering
\begin{subfigure}{0.99\columnwidth}

    \includegraphics[trim={0cm 0cm 0cm 0cm},clip,width=\linewidth]{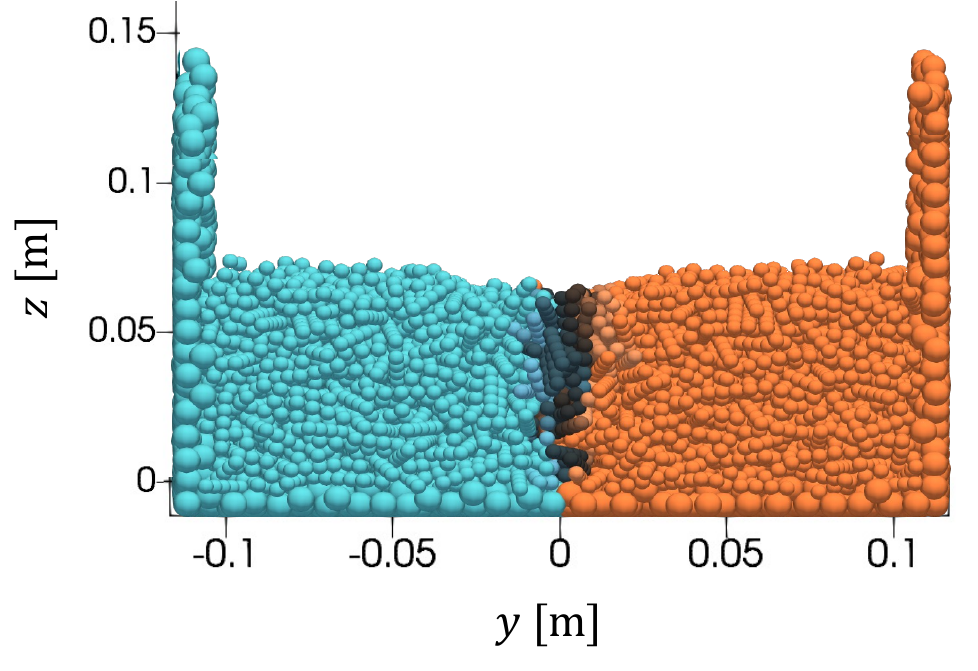}
    \hspace*{1cm} (a)
\end{subfigure}

\begin{subfigure}{0.9\columnwidth}
    \includegraphics[trim={0cm 0cm 0cm 0cm},clip,width=\linewidth]{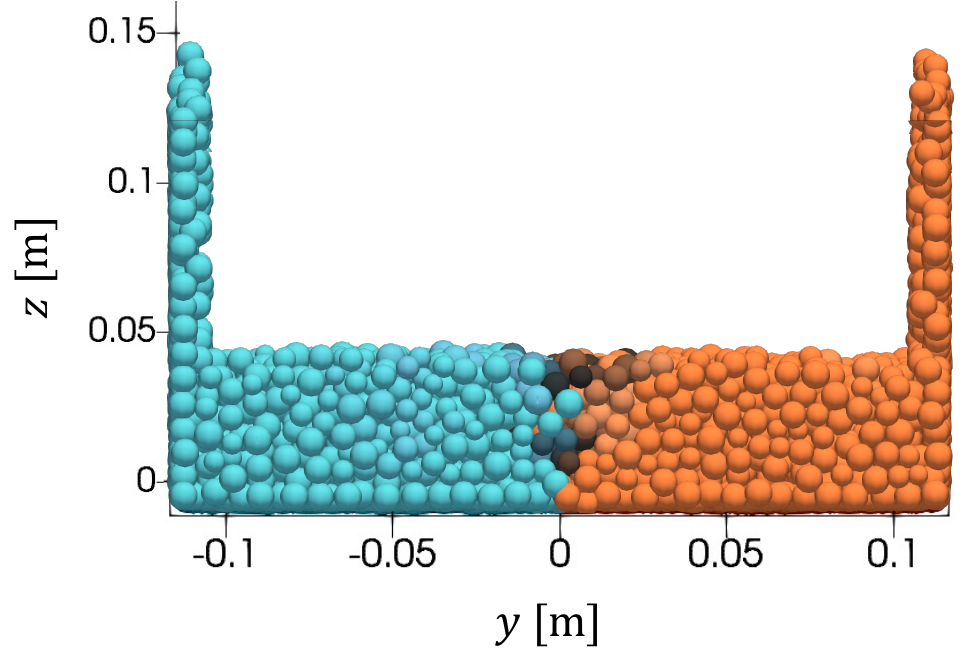}
    \hspace*{1cm} (b)
\end{subfigure}

\hspace{1cm}
\begin{subfigure}{0.8\columnwidth}
    \includegraphics[trim={0cm 0cm 0cm 0cm},clip,width=\linewidth]{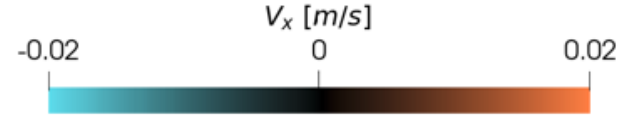}
\end{subfigure}
\caption{The shear cell contains (a) elongated particles with a depression above the shear band and (b) spherical particles with a flat surface.}
    \label{fig:LSC_simulation}
\end{figure}

\autoref{fig:LSC}(a) sketches the system:
\begin{figure}[!htbp]
\centering
\begin{subfigure}{0.99\columnwidth}
    \includegraphics[trim={0cm 0cm 0cm 0cm},clip,width=\linewidth]{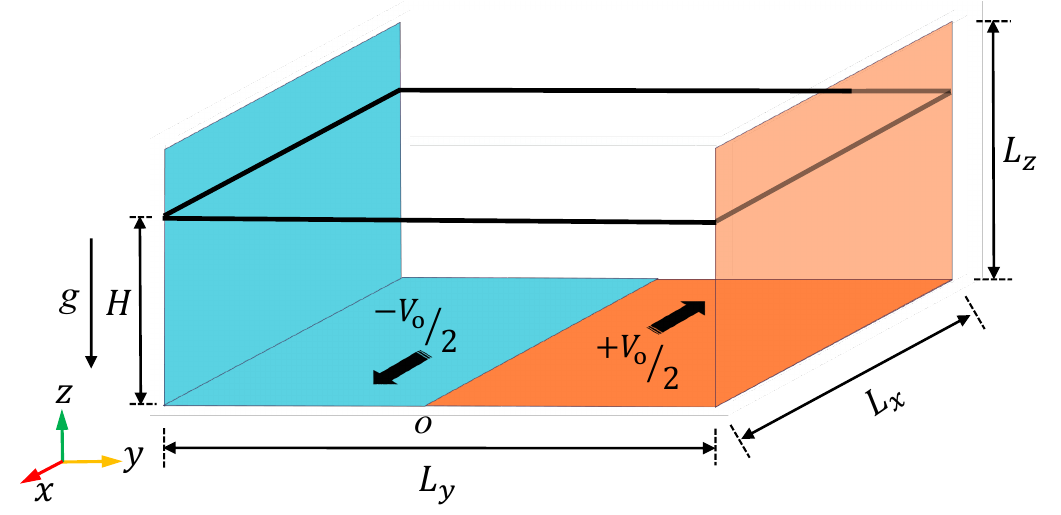}
    \subcaption{}
\end{subfigure}
\begin{subfigure}{0.9\columnwidth}
    \includegraphics[trim={0cm 6cm 0cm 0cm},clip,width=\linewidth]{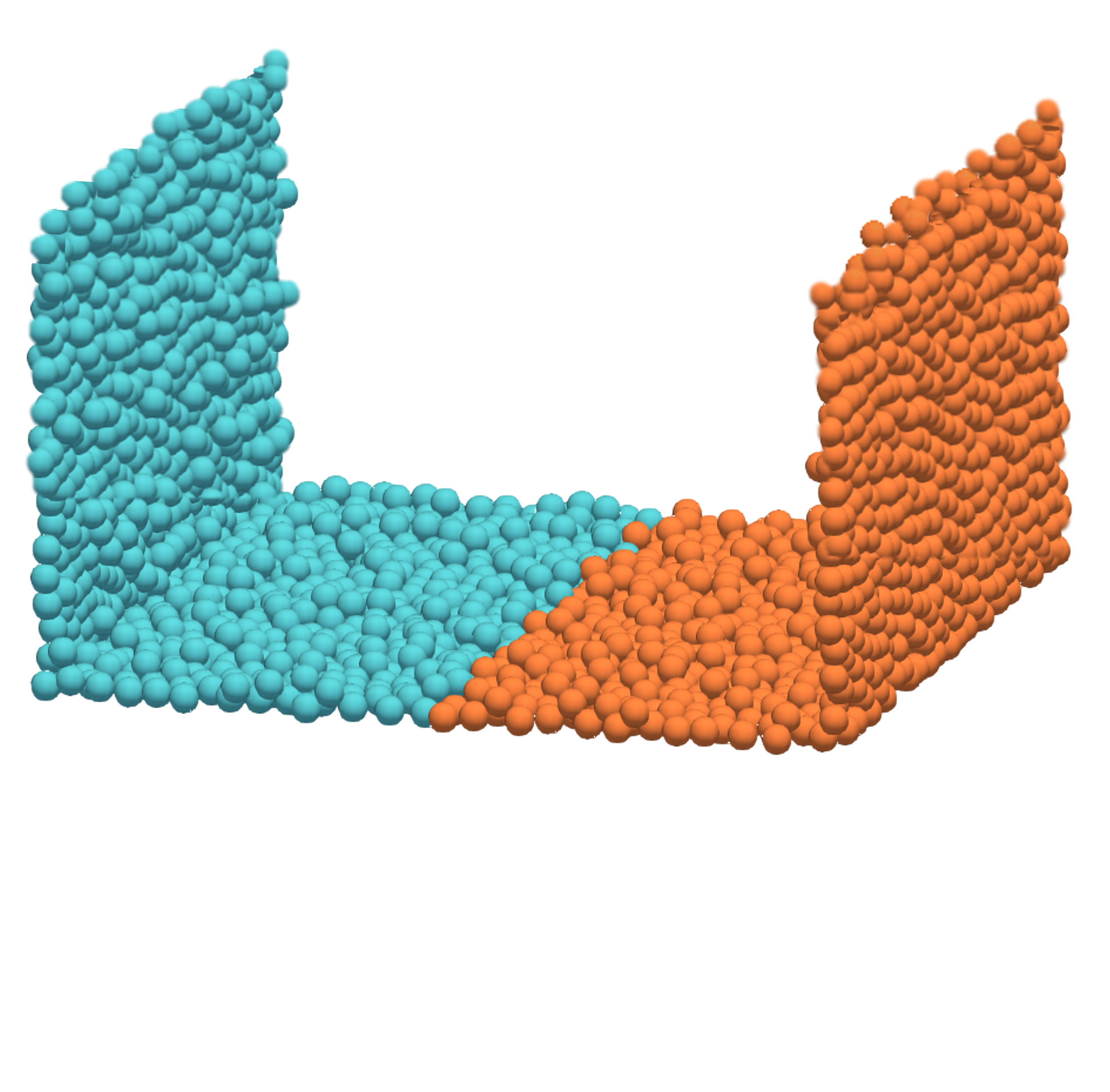}
    \subcaption{}
\end{subfigure}
\caption{(a) Sketch of the linear split-bottom shear cell (LSC): $y=0$ indicates the split position, the shear velocity is $V_0$, and $g$ shows the direction of gravity. (b) Spheres are used to model the walls of the shear cell.}
    \label{fig:LSC}
\end{figure}
The LSC consists of two L-shaped walls that move at velocity $V_0/2$ in opposite directions along the $x$-axis. The position of the slit, $y=0$, defines the location where the profiles move relative to each other. Gravity $g$ acts in the negative $z$-direction. The system parameters are chosen to match those used in the cylindrical split-bottom experiments by Fischer et al. \cite{fischer2016heaping}, where the shear velocity $V_0=0.038\,\text{m/s}$ corresponds to 3 rpm for a rotating disk of radius $R_\text{disk} = 118,\text{mm}$. The LSC is periodic in the $x$-direction. To avoid slip at the boundaries, the walls are modeled using spheres of diameter $d_{p} =8.55\,\text{mm}$, see \autoref{fig:LSC}(b). The size of the LSC is $(L_{x}, L_{y}, L_{z}) = (25, 25, 20)d_{p}$, and the filling height is denoted by $H$. When the granular material is sheared, a shear band emerges from the split position, extending in both the $y$ and $z$ directions, as shown by the regions of velocity gradient in \autoref{fig:LSC_simulation}. We use the same numerical setup as in our previous work ~\cite{rahim2024alignment, rahim2025impact}, where surface profile evolution was studied for particles with varying $\text{AR}$ and $\mu$. Here, we analyze the spatial variation of pressure and stress fields within the shear band and in the surrounding bulk. All profiles shown in this work are averaged along the homogeneous $x$-direction and plotted in the $y$–$z$ plane.

\subsection{Particle shape}
\begin{figure}[htbp]
    \centering
    \includegraphics[trim={0cm 0cm 0cm 0cm},clip,width=0.8\columnwidth]{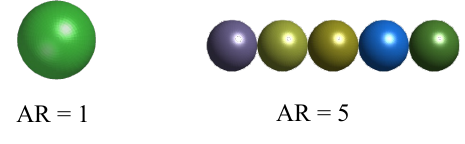}
    \caption{Particle shapes used in simulations: spheres ($\text{AR} = 1$) and elongated particles ($\text{AR} = 5$).}
    \label{fig:particle_shape}
\end{figure}

The aspect ratio, defined as the length-to-diameter ratio of the elongated particles, is used to distinguish particle shapes. We consider two particle shapes: spheres ($\text{AR} = 1$) and elongated particles ($\text{AR} = 5$), as illustrated in \autoref{fig:particle_shape}. 
The elongated particle is modeled using the multi-sphere method \cite{Buchholtz:1993} as a linear chain of five identical spheres. This approach simplifies contact detection, since all contacts are handled as sphere–sphere interactions \cite{abou2004three, kodam2009force, cabiscol2018calibration}. To introduce size polydispersity, we sample $N = 4,500$ particle diameters from a uniform distribution with a mean of $\langle d \rangle = 7.6,\text{mm}$ and ranging over $\langle d \rangle$ $\pm 20\%$. For elongated particles, the diameters of the constituent spheres are scaled by a factor of $\text{AR}^{1/3}$ to keep the total particle volume the same as that of spherical particles. The particle count and size distribution are consistent with previous experiments \cite{ries2007shear}.

\subsection{Contact model and material parameters}
The visco-elastic Hertz-Mindlin contact model \cite{BSHP:1996, mindlin1949compliance} 
is used to describe the force between spheres in contact. The normal component reads \cite{BSHP:1996}
\begin{equation}
    \vec{F}_n = \min\left(0, -k\xi^{3/2} - \frac{3}{2}A_n k\sqrt{\xi}\dot{\xi}\right) \vec{e}_n\,,
\end{equation}
where $\xi = R_i + R_j - \left|\vec{r}_i-\vec{r}_j\right|$ is the compression of spheres $i$ and $j$ of radii $R_i$ and $R_j$ at positions $\vec{r}_i$ and $\vec{r}_j$, $\vec{e}_n = (\vec{r}_i-\vec{r}_j)/\left|\vec{r}_i-\vec{r}_j\right|$ is the normal unit vector. The normal dissipative parameter $A_n = 6\times 10^{-5}$s corresponds to the coefficient of restitution $0.4$ for a particle of radius $2.5\,\text{mm}$ and elastic modulus $ E = 10\,\text{MPa}$, impacting at a velocity $2\,\text{m/s}$ \cite{muller2011collision}. 
The effective stiffness of the Hertzian contact model is
\begin{equation}
    k = \frac{4}{3} \, E^* \, \sqrt{R^*} 
    \label{eq:Hertz_rho}
\end{equation}
with the effective radius $R^*$ and the effective elastic modulus
\begin{equation}
     E^* = \left(\frac{1-\nu_i^2}{E_i} + \frac{1-\nu_j^2}{E_j}\right)^{-1}\,,
\end{equation}
where $E_{i}$ is the elastic modulus and $\nu_{i}$ is its Poisson ratio of the material of particle $i$. 

For the tangential viscoelastic force, we assume the no-slip expression by Mindlin \cite{mindlin1949compliance} for the elastic part and Parteli and P\"oschel \cite{parteli2016particle} for the tangential dissipative constant $A_t \approx 2 A_n E^*$. The force is limited by the Coulomb criterion:
\begin{equation}
    \vec{F}_t = -\min \left[ \mu\left|\vec{F}_n\right|,  \int 8 G^{*}\sqrt{R^* \xi} \,ds 
    + A_t  \sqrt{R^* \xi} v_t \right] \vec{e}_t\,,
\end{equation}
with the friction coefficient, $\mu$, and the effective shear modulus 
\begin{equation}
    G^*=\left(\frac{2-\nu_i}{G_i} + \frac{2-\nu_j}{G_j}\right)^{-1},
\end{equation}
 which, for identical materials, simplifies to 
 \begin{equation}
     G^*=\frac{G}{2\left(2-\nu\right)}\,.
 \end{equation} 
 The integral is performed over the displacement of the colliding particles at the point of contact, for the total duration of the contact \cite{parteli2016particle}.
The material parameters corresponding to wooden pegs are given in \autoref{tab:material_parameters} \cite{fischer2016heaping}.
\begin{table}[htb]
\caption{DEM simulation parameters}
\label{tab:material_parameters}
\begin{center}
\begin{tabular}{l@{\quad}l@{\quad}ll}
\hline
variable & unit& value\\
\hline
elastic modulus ($E$)  & MPa & 10\\
sliding friction coefficient ($\mu$)  & - & 0.01-0.8\\
Poisson's ratio ($\nu$)  & -& 0.35\\
particle density ($\rho$) & kg/m$^3$& 850\\
\hline                
\end{tabular}
\end{center}
\end{table}

We use the material density and friction coefficient of wooden pegs \cite{fischer2016heaping}. For acceptable computer time, the Young’s modulus used in the simulations is smaller than the Young's modulus of the wooden pegs. With the chosen value, the maximum particle compression is less than 2\% of the particle diameter, which is within the acceptable limits for DEM simulations \cite{luding2008introduction, thornton2000numerical, cundall1979discrete}.

\section{Coarse-graining of stationary fields}
We aim to derive macroscopic fields—such as strain rate, velocity profile, shear band width, packing density, and stress—that correspond to the features of the shear band and macromechanical stress analysis, based on the given micromechanical properties. To achieve this, we use a coarse-graining method to calculate these macro-parameters, utilizing precise sphere overlap volumes and mesh elements as outlined by Strobl et al. \cite{strobl2016exact}.

We simulated the system for 500\,s in real-time. The shear displacement $\lambda = V_{0}\,t$ was used to ensure the steady state (where $t$ is the simulation time). The kinetic energy and average contact number converge at $\lambda \approx 47 \, d_p$ (10 s), while a stable shear band and flow profile required $\lambda \approx 280 \, d_p$ \cite{singh2024shear, ries2007shear}. The stress tensor is computed locally at each spatial point using the standard DEM formulation based on contact forces and branch vectors \cite{weinhart2016influence, singh2015role}. The fields are averaged over the (periodic) $x$-direction in the time interval $t\in (400, 500)\,\text{s}$ when the system adopted its stationary state. From these fields, we obtained the stationary quantities studied here.

\section{Results and discussions}

\subsection{Pressure distribution in the YZ plane}
To understand how pressure varies within the sheared granular material, we analyze its distribution in the $ yz$ plane, averaged along the $x$ direction. The stress tensor is computed from contact forces and branch vectors following the standard DEM formulation \cite{weinhart2016influence, singh2015role}. The pressure $P$ is defined as the average of the normal stress components, which corresponds to the trace of the stress tensor \cite{singh2015role}:
\begin{equation}
    P = \frac{1}{3}(\sigma_{xx} + \sigma_{yy} + \sigma_{zz}),
\end{equation}

where $\sigma_{xx}$, $\sigma_{yy}$, and $\sigma_{zz}$ are the normal stresses in shear, lateral, and vertical directions, respectively. These stresses are extracted from the shear band region, which has reached the critical state.

\begin{figure}[!htbp]
\centering
\includegraphics[trim={0cm 0cm 0cm 0cm},clip,width=0.99\columnwidth]{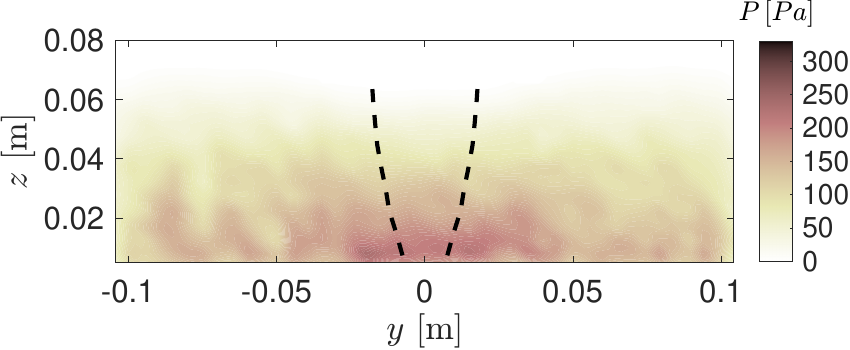}
\caption{Pressure field in the $yz$-plane, averaged along the $x$-axis. The dashed lines indicate the shear band region, calculated from the dimensionless velocity profile.}
\label{fig:press_in_YZ_plane} 
\end{figure}

\autoref{fig:press_in_YZ_plane} shows the resulting pressure field in the $yz$ plane, averaged along the $x$-axis. The dashed lines indicate the shear band boundaries. The width of the shear band is determined by fitting the normalized velocity profile in the $x$ direction as a function of $y$ to an error function \cite{dijksman2010granular, ries2007shear, depken2006continuum}. The shear-band width is taken as the characteristic length scale from this fit. The pressure increases with depth (towards lower $z$) because of the gravitational load from the particles above.

\subsubsection{Pressure profile along the \texorpdfstring{$z$}{z}-axis}
\begin{figure}[htbp]
\centering
\includegraphics[trim={0cm 0cm 0cm 0cm},clip,width=0.99\columnwidth]{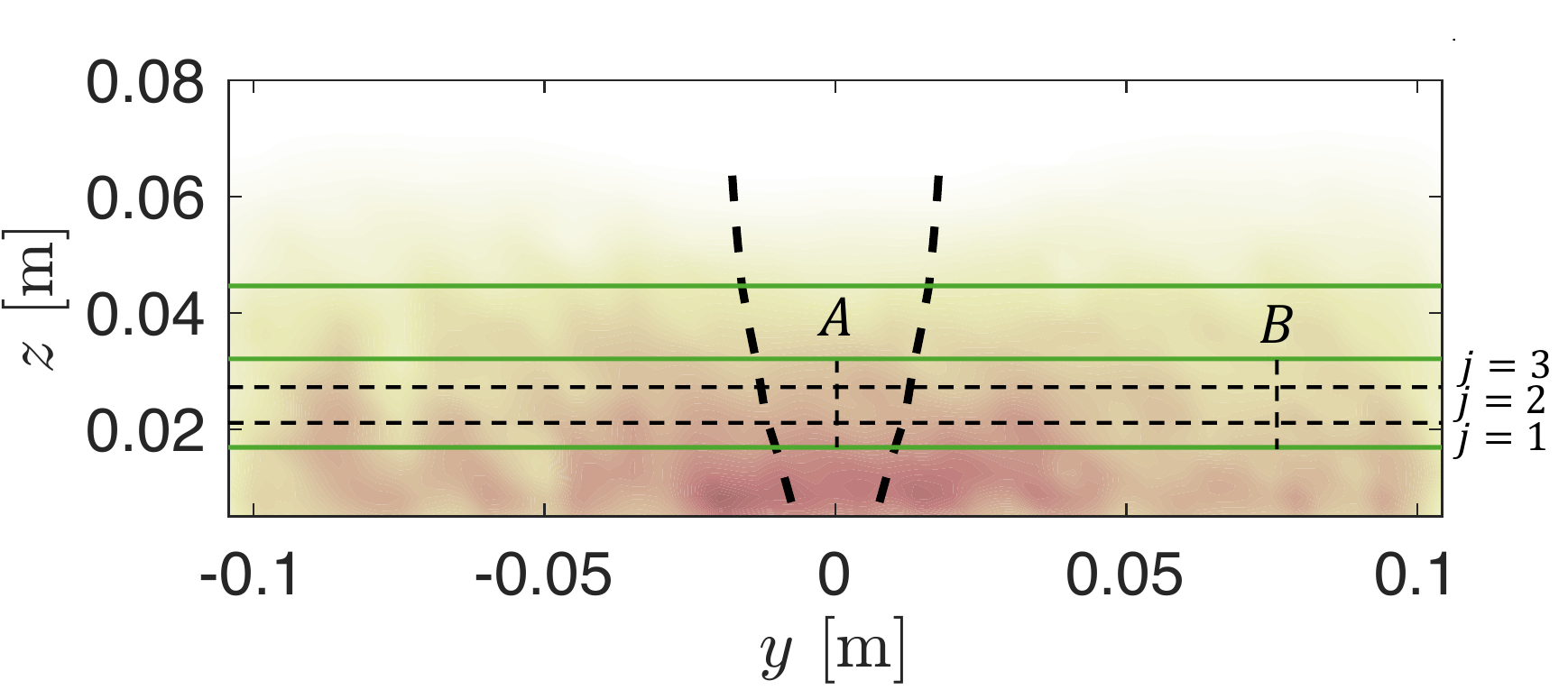}    
\caption{Sampling positions and $z$-slices used for pressure analysis. Solid lines show the three bulk layers; dashed lines show the three slices within the middle layer ($j=1,2,3$). $A$ and $B$ denote sampling positions inside and outside the shear band, respectively.}
\label{fig:press_layering}
\end{figure}

To study how pressure varies with height ($z$) at different lateral ($y$) positions, the bulk material is divided into three horizontal layers along the $z$-axis (indicated by the solid lines in \autoref{fig:press_layering}). For the analysis, we consider only the middle layer, because the top and bottom layers are influenced by boundaries. This middle layer is then divided into three horizontal slices, labeled $j = 1, 2, 3$ from bottom to top (dashed lines in \autoref{fig:press_layering}).


Pressure is sampled at two fixed positions: position $A$ at the center of the shear band ($y = 0$), and position $B$ in the bulk ($y = 0.08\, \text{m}$), outside the shear band. Position $B$ is used as the bulk reference. For each slice $j$, the average pressure at position $Q \in \{A, B\}$ is computed as:

\begin{equation}
\langle P \rangle^{Q}_j = \frac{1}{N^{Q}_j} \sum_{i=1}^{N^{Q}_j} P^{Q}_i,
\end{equation}
where $P^{Q}_i$ is the coarse-grained pressure in the $i$-th cell of slice $j$ at position $Q$, and $N^{Q}_j$ is the number of such cells in that slice.

\begin{figure}[htbp]
\centering
\includegraphics[trim={0cm 0cm 0cm 0cm},clip,width=0.99\columnwidth]{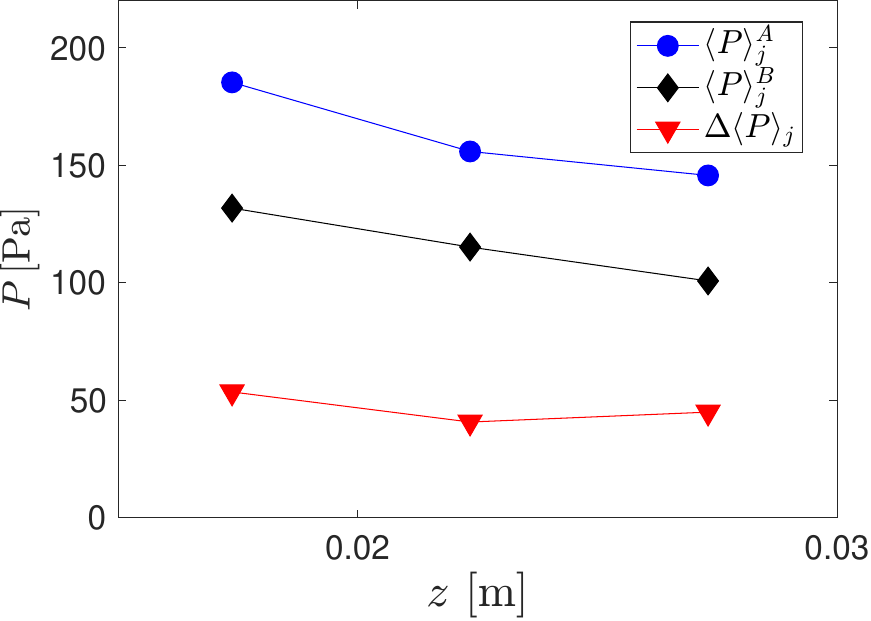} 


\caption{Pressure as a function of height ($z$) at two lateral positions: inside the shear band ($A$ at $y = 0$) and outside it ($B$ at $y = 0.08$).}
\label{fig:press_vs_z}
\end{figure}

\autoref{fig:press_vs_z} shows $\langle P\rangle^{Q}_j$ as a function of the slice height $z_j$. Each data point corresponds to the pressure averaged within slice $j$ at position $A$ or $B$, i.e., $\langle P\rangle_j^A$ and $\langle P\rangle_j^B$. The pressure decreases with increasing $z$, indicating the decreasing weight of the overlying particles. At all $z$, position $A$ exhibits higher pressure than position $B$, resulting in a positive pressure difference ($\Delta P_j = \langle P \rangle_j^A - \langle P \rangle_j^B > 0$). This shows that the pressure at the center of the shear band is consistently higher than the pressure outside it.

\subsubsection{Pressure profile along the \texorpdfstring{$y$}{y}-axis}
We compute the pressure averaged over height ($z$) within the middle layer (slices $j = 1$ to $3$) to determine whether the high pressure in the shear band is localized or extends away from $y = 0$. The resulting pressure profile, $\langle P \rangle_z(y)$, is:
\begin{equation}
    \langle P \rangle_{z} (y) = \frac{1}{N} \sum_{i=1}^{N} P_{i}
\end{equation}
where $P_i$ is the pressure in the $i$-th coarse-grained cell at position $y$, and $N$ is the total number of such cells in the middle layer.

\begin{figure}[!htbp]
\centering
\begin{subfigure}{0.99\columnwidth}
    \includegraphics[trim={0cm 0cm 0cm 0cm},clip,width=\columnwidth]{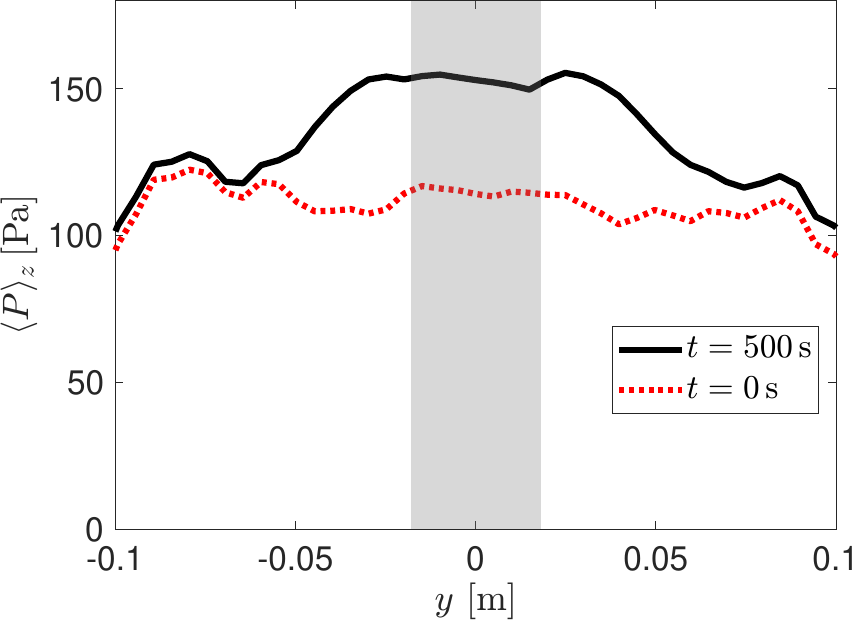}
    \hspace*{1cm} (a)
\end{subfigure}
\begin{subfigure}{0.99\columnwidth}
    \includegraphics[trim={0cm 0cm 0cm 0cm},clip,width=\columnwidth]{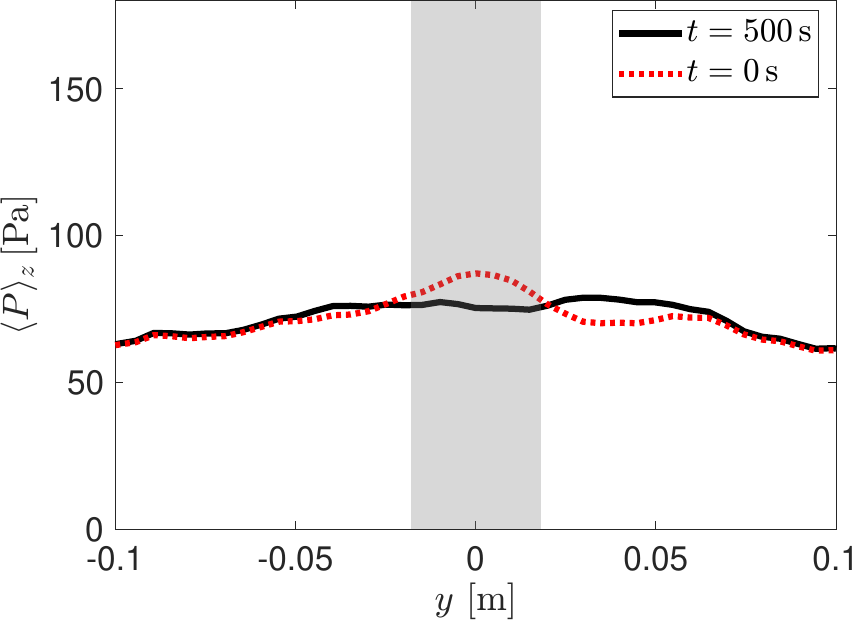}
    \hspace*{1cm} (b)
\end{subfigure}
\caption{Average pressure along the $y$-axis at $t = 0$~s and $t = 500$~s for (a) elongated particles showing a localized pressure peak under shear, and (b) spherical particles, showing uniform pressure across $y$. The gray-shaded region indicates the width of the shear band averaged along $z$.}
\label{fig:press_vs_y} 
\end{figure}

\autoref{fig:press_vs_y} shows $ \langle P \rangle_{z}(y)$, both before shearing ($t = 0$~s) and at steady state ($t = 500$~s) for both elongated and spherical particles. The gray shaded region indicates the width of the shear band, averaged over $z$. 

For elongated particles \autoref{fig:press_vs_y}(a), the pressure is nearly uniform across $y$ at $t = 0$. At steady state ($t = 500$~s), however, the pressure at the center of the shear band is larger than the pressure outside it. This indicates that the pressure difference is shear-induced and spatially localized. Thus, position $B$, chosen outside the shear band, serves as a reliable reference for low pressure. Spherical particles (\autoref{fig:press_vs_y}(b)) exhibit nearly uniform pressure across $y$ at both $t = 0$ and $t = 500$~s. The isotropic pressure is due to their symmetric shape, which allows easy rearrangement under shear and prevents stress localization.

In the following sections, we apply the same $z$-averaging approach to extract other coarse-grained fields, such as packing density and stress components, at the two fixed lateral positions $A$ and $B$. For each position $Q \in \{A, B\}$, the coarse-grained field value $F$ is averaged over the middle layer (slices $j = 1$ to $3$) as:
\begin{equation}
\langle F \rangle_{z}^Q = \frac{1}{N^{Q}} \sum_{i=1}^{N^{Q}} F_i^Q,
\label{eq:field_avg_point}
\end{equation}

where $F_i^Q$ is the value of the field in the $i$-th coarse-grained cell at position $Q$, and $N^{Q}$ is the number of such cells in the middle layer.

\subsubsection{Time evolution of pressure and normal stress components}
\begin{figure}[!htbp]
\centering
\begin{subfigure}{0.99\columnwidth}
    \includegraphics[trim={0cm 0cm 0cm 0cm},clip,width=\columnwidth]{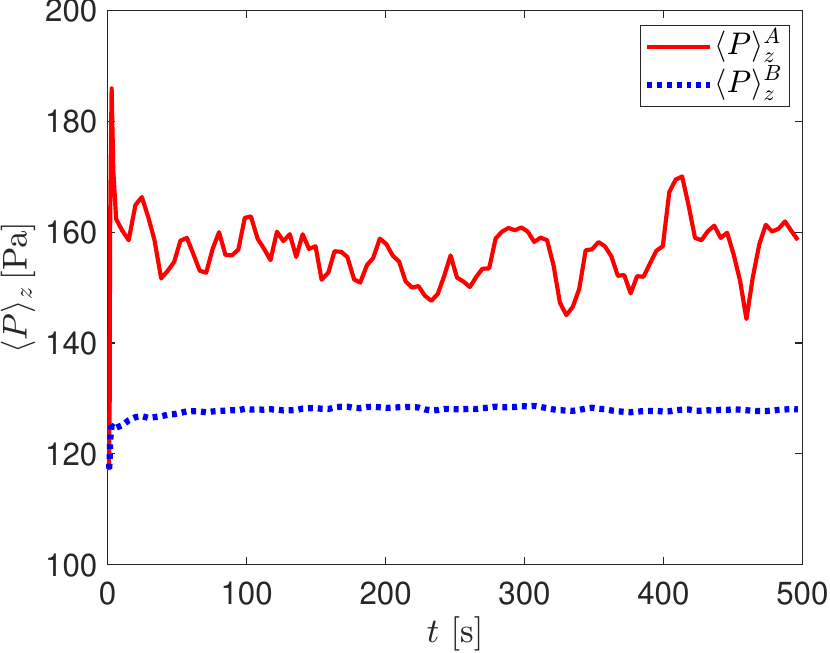}
    \hspace*{1cm} (a)
\end{subfigure}
\begin{subfigure}{0.99\columnwidth}
    \includegraphics[trim={0cm 0cm 0cm 0cm},clip,width=\columnwidth]{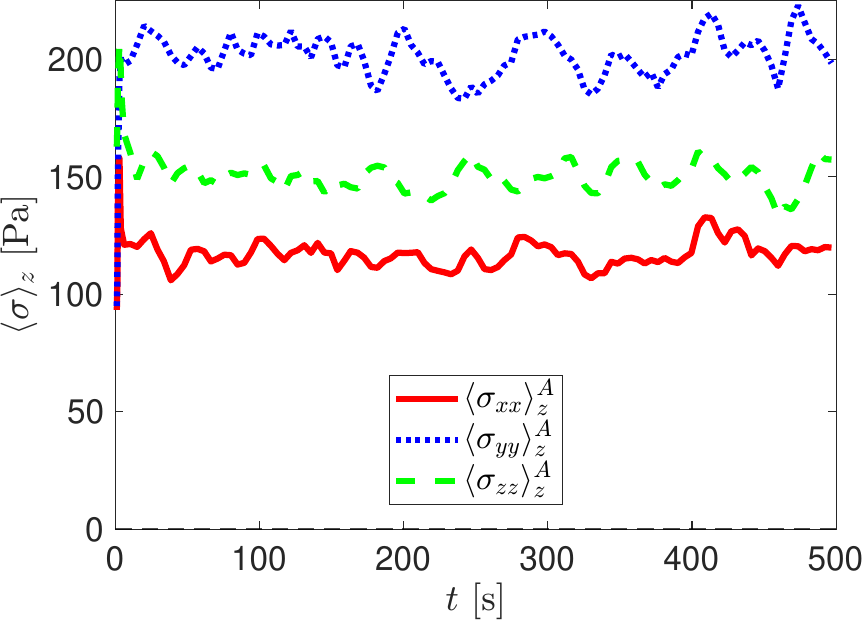}
    \hspace*{1cm} (b)
\end{subfigure}
\caption{Pressure and stress evolution for elongated particles. (a) $\langle P \rangle_z$ as a function of time at positions $A$ and $B$, showing the development of a pressure difference under shear. (b) Normal stress components at position $A$ for elongated particles, where $\sigma_{yy}$ and $\sigma_{zz}$ are consistently higher than $\sigma_{xx}$.}
\label{fig:AR5_pressure_time} 
\end{figure}

We compute the $z$-averaged pressure at each time step at positions $A$ and $B$, denoted by $\langle P \rangle_z^A$ and $\langle P \rangle_z^B$. Their time evolution is shown in \autoref{fig:AR5_pressure_time}(a).

At $t = 0$, the pressure is nearly equal at both positions. As shear begins, both $\langle P \rangle_z^A$ and $\langle P \rangle_z^B$ increase and then reach a steady value. In the steady state, $\langle P\rangle_z^A$ is consistently higher than $\langle P\rangle_z^B$, showing that shear produces a localized pressure increase inside the shear band.

To understand this pressure difference, we examine the normal stress components at position $A$: $\langle \sigma_{xx} \rangle_z^A$, $\langle \sigma_{yy} \rangle_z^A$, and $\langle \sigma_{zz} \rangle_z^A$, representing normal stresses in the shear, lateral, and vertical directions, respectively. As shown in \autoref{fig:AR5_pressure_time}(b), all three components increase as shear begins and reach a steady state. $\sigma_{xx}$ stabilizes at a lower value due to particle alignment in the shear direction, which decreases resistance to flow. $\sigma_{yy}$ increases due to wall confinement and increased contact density. $\sigma_{zz}$ increases initially but relaxes to its pre-shear value as dilatancy reduces vertical loading.

\begin{figure}[!htbp]
\centering
\begin{subfigure}{0.99\columnwidth}
    \includegraphics[trim={0cm 0cm 0cm 0cm},clip,width=\columnwidth]{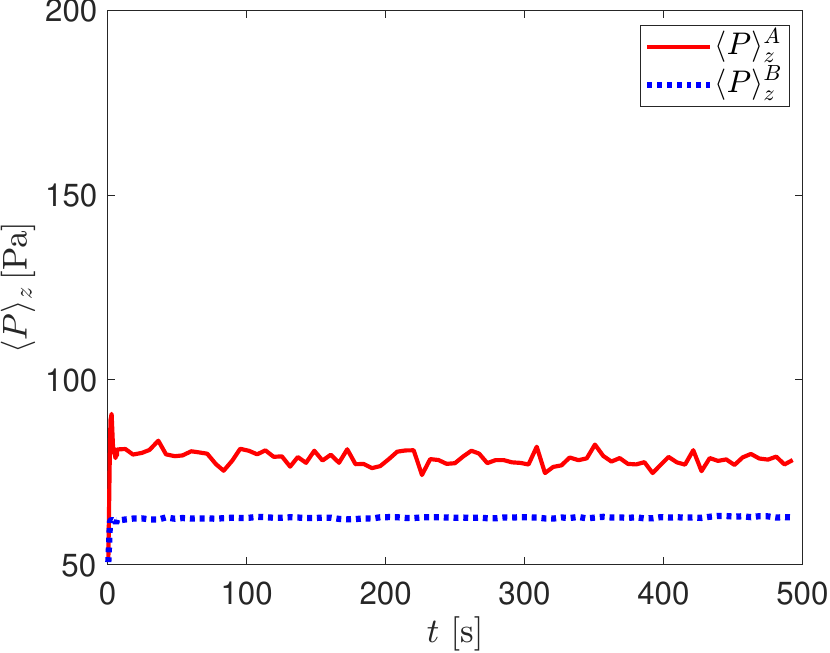}
    \hspace*{1cm} (a)
\end{subfigure}
\begin{subfigure}{0.99\columnwidth}
    \includegraphics[trim={0cm 0cm 0cm 0cm},clip,width=\columnwidth]{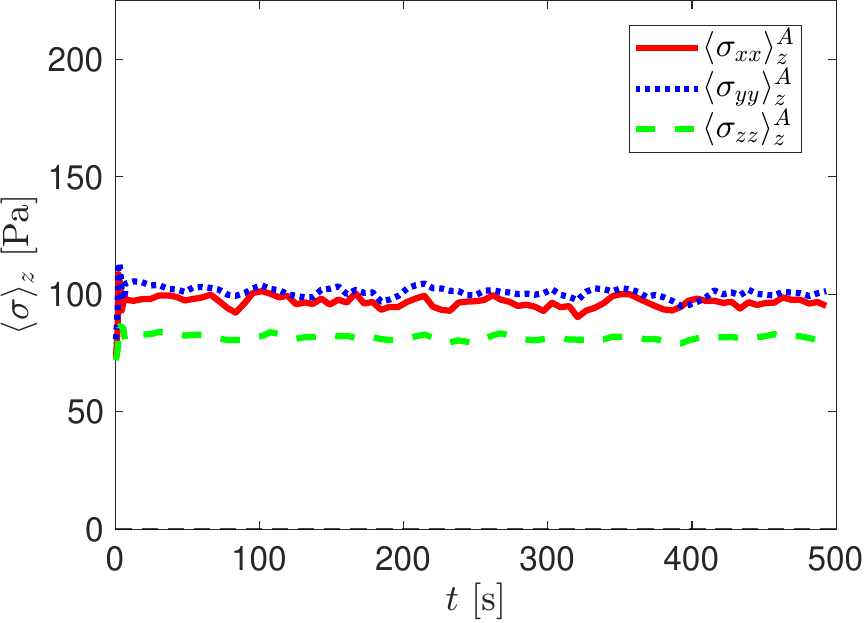}
    \hspace*{1cm} (b)
\end{subfigure}
\caption{Pressure and stress evolution for spherical particles. (a) $\langle P \rangle_z$ as a function of simulation time ($t$) at positions $A$ and $B$, showing a small difference, $\langle P\rangle_z^A$ slightly higher than $\langle P\rangle_z^B$. (b) Normal stress components at position $A$, showing minute differences and are nearly isotropic.}
\label{fig:AR1_pressure_time} 
\end{figure}

\autoref{fig:AR1_pressure_time}(a) shows the corresponding pressure evolution for spherical particles. Initially, both positions exhibit similar pressure. As shear begins, pressure increases slightly and stabilizes. In the steady state, $\langle P\rangle_z^A$ is slightly larger than $\langle P\rangle_z^B$, but the difference is much smaller than that of elongated particles, indicating a more uniform stress state. \autoref{fig:AR1_pressure_time}(b) shows the corresponding normal stresses at position $A$, all of which increase slightly and then stabilize. These small differences between the stress components indicate an almost isotropic stress distribution. Although friction can induce anisotropy, spherical particles rearrange easily, minimizing this effect.

Since spherical particles do not exhibit a pronounced pressure difference between the shear band and the surrounding region, the following sections focus on elongated particles to investigate the origin of the observed pressure anisotropy.

\subsection{Packing density}
We analyze the packing density ($\phi$) at positions $A$ and $B$ to understand why pressure is higher inside the shear band. The average packing density $\langle \phi \rangle_z^Q$ is obtained by averaging $\phi$ over the three $z$-slices in the middle layer at each position $Q \in \{A, B\}$.

\begin{figure}[!htbp]
\centering
\includegraphics[trim={0cm 0cm 0cm 0cm},clip,width=0.99\columnwidth]{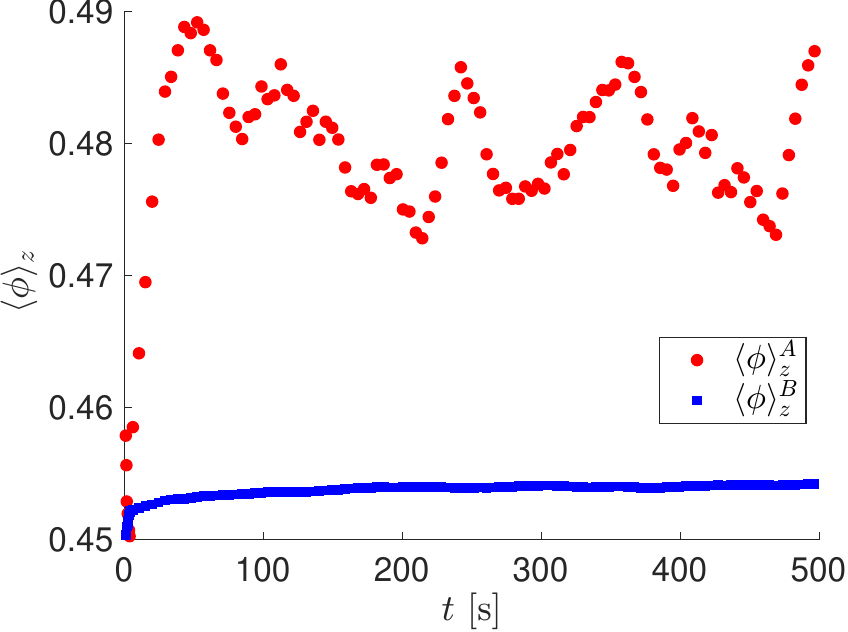}
\caption{Packing density $\langle \phi \rangle_z$ over time at positions $A$ and $B$ for elongated particles. $\langle \phi \rangle_z^A$ is consistently larger than $\langle \phi \rangle_z^B$ in the steady state.}
\label{fig:pd_time} 
\end{figure}

\autoref{fig:pd_time} shows the time evolution of $\langle \phi \rangle_z^A$ and $\langle \phi \rangle_z^B$. Both positions exhibit similar packing density at $t = 0$. After shear begins, the packing density increases at both positions and stabilizes. In the steady state, $\langle \phi \rangle_z^A$ is consistently higher than $\langle \phi \rangle_z^B$. This denser packing at position $A$ indicates a larger number of particle contacts, which contributes to the higher pressure inside the shear band.

\subsection{Particle alignment}
Shear reorients elongated particles, influencing both packing density and stress distribution \cite{rahim2024alignment, borzsonyi2012orientational}. 
To investigate the origin of the higher packing density inside the shear band, we compare particle alignment at positions $A$ and $B$. We define an angle $\theta_x$ to measure how much a particle deviates from the shear direction, as the angle between the particle’s major axis (unit vector $\vec{P}_v$) and the shear direction (unit vector $\vec{X}$):

\begin{equation}\label{dotprod}
   \theta_{x} = \frac{\pi}{2} - \left|\frac{\pi}{2} - \arccos \left(\frac{\Vec{P_{v}} \cdot \Vec{X}}{|\Vec{P_{v}}||\Vec{X}|}\right) \right|\,,
\end{equation}

$\theta_x = 0$ indicates perfect alignment with the shear direction, while $\theta_x = \pi/2$ corresponds to complete misalignment. The average alignment at each position $Q \in \{A,B\}$ is computed by averaging $\theta_x$ over all particles located in the three $z$-slices of the middle layer:

\begin{equation} \langle \theta_x \rangle_z^Q = \frac{1}{N} \sum_{i=1}^{N} \theta_{x,i}^Q\, ,
\end{equation}
where $N$ is the number of particles located at position $Q$ within the three horizontal slices of the middle layer.

\begin{figure}[!htbp]
\centering
\includegraphics[trim={0cm 0cm 0cm 0cm},clip,width=0.99\columnwidth]{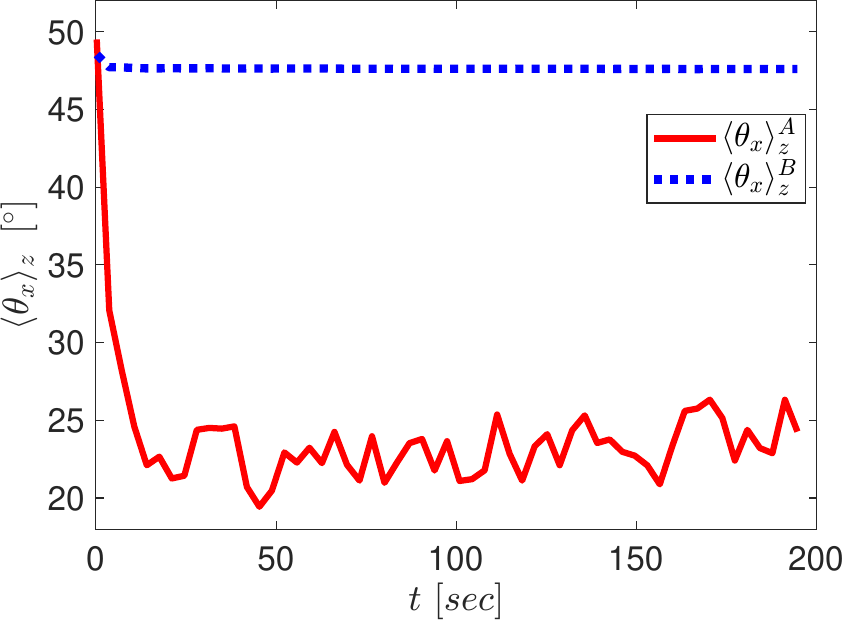}
    \caption{$\langle \theta_x \rangle_z$ as a function of time ($t$) at positions $A$ and $B$. Smaller values at $A$ indicate a more pronounced alignment with the shear direction.}
    \label{fig:alignment_time}
\end{figure}

\autoref{fig:alignment_time} shows how $\langle \theta_x \rangle_z$ evolves over time at positions $A$ and $B$. At $t = 0$, particles are randomly oriented, and $\langle \theta_x \rangle_z$ is large at both positions. As shear progresses, $\langle \theta_x \rangle_z^A$ decreases rapidly and then fluctuates around $20^\circ\text{–}25^\circ$, indicating strong alignment with the shear direction. At position $B$, $\langle \theta_x \rangle_z^B$ changes only slightly and remains close to its initial value. This stronger alignment at $A$ decreases interparticle gaps, leading to denser packing and therefore large pressure inside the shear band.


\subsection{Influence of friction}
We study the effect of friction ($\mu$) on the pressure distribution by simulating elongated particles ($\text{AR}=5$) and spherical particles ($\text{AR}=1$) with $\mu \in [0.01, 0.8]$.

\begin{figure}[!htbp]
\centering
\begin{subfigure}{0.80\columnwidth}
    \includegraphics[width=\linewidth]{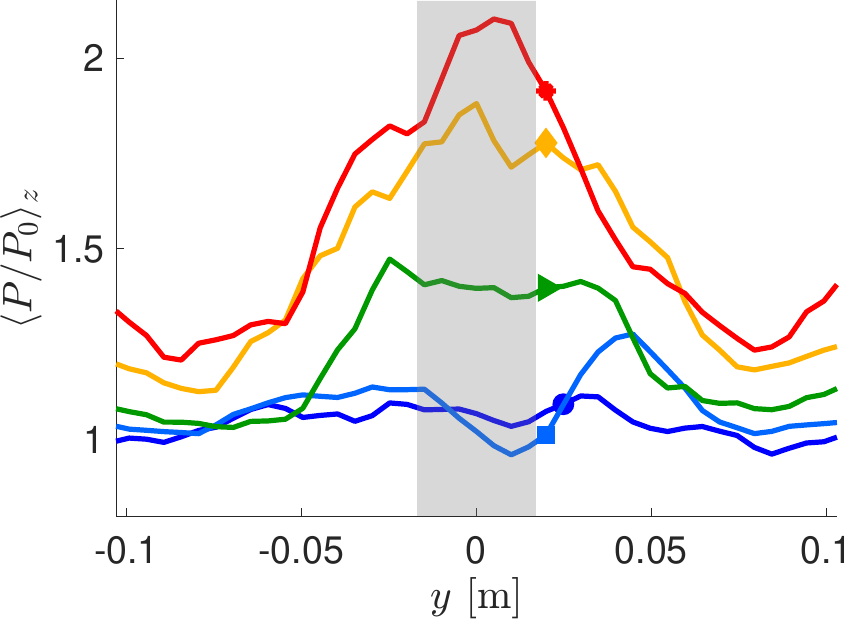}
    \hspace*{1cm} (a)
\end{subfigure}
\hfill
\begin{subfigure}{0.80\columnwidth}
    \includegraphics[width=\linewidth]{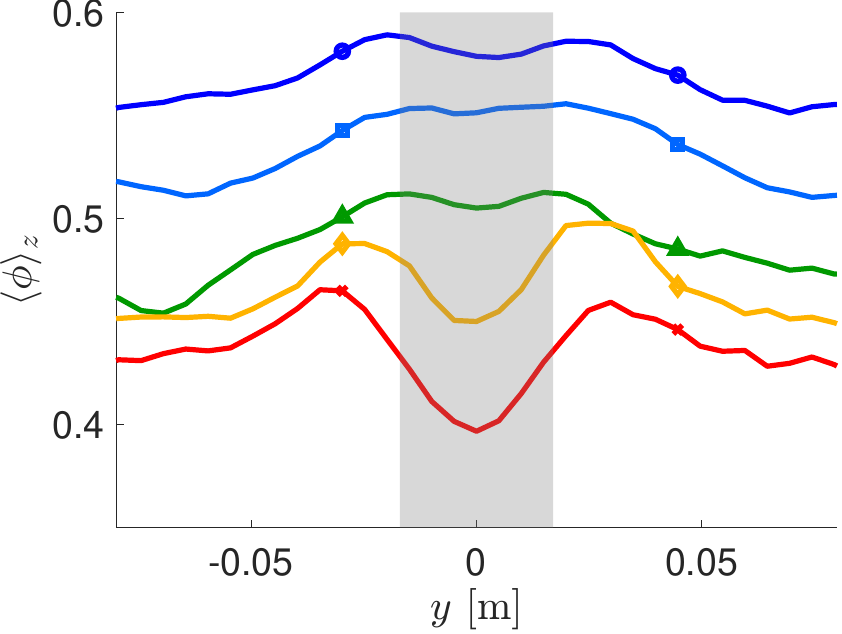}
    \hspace*{1cm} (b)
\end{subfigure}

\begin{subfigure}{0.80\columnwidth}
    \includegraphics[width=\linewidth]{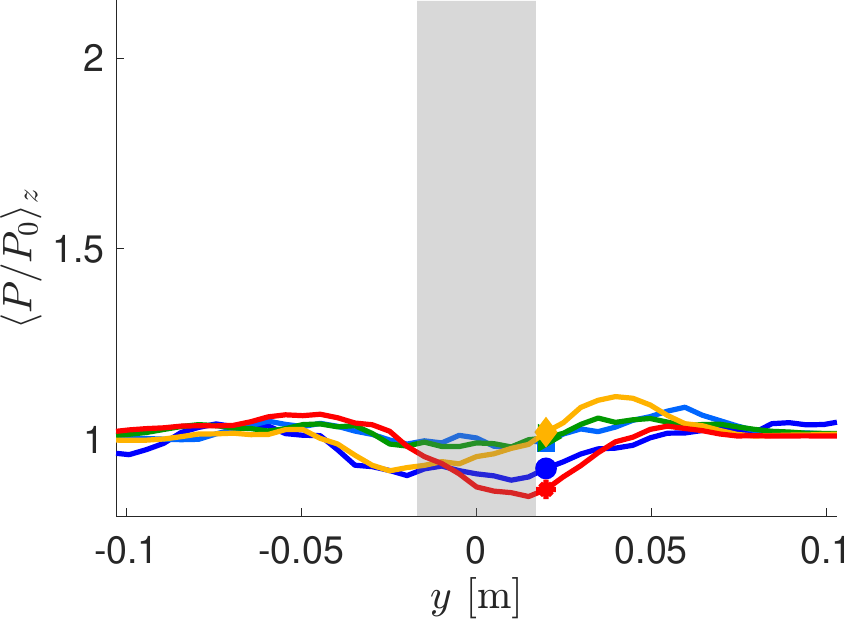}
    \hspace*{1cm} (c)
\end{subfigure}
\begin{subfigure}{0.85\columnwidth}
    \includegraphics[width=\linewidth]{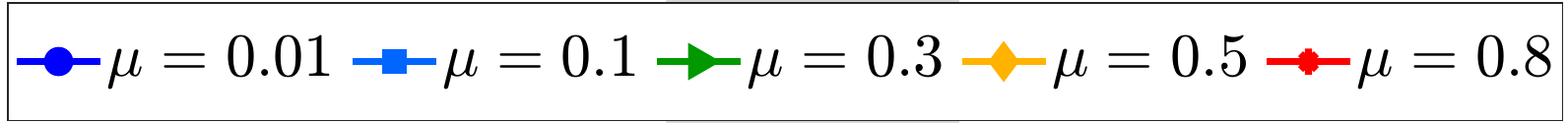}
\end{subfigure}
\caption{$\langle P/P_0 \rangle_z$ as a function of $y$ for elongated particles ($\text{AR}=5$) at various $\mu$.
(b) $\langle \phi \rangle_z$ for elongated particles at the same $\mu$, showing shear-induced dilation at higher $\mu$.
(c) $\langle P/P_0 \rangle_z$ for spherical particles ($\text{AR}=1$), showing no localization and nearly uniform pressure for all $\mu$.
The grey shaded region indicates the shear band width averaged along $z$.}
\label{fig:AR_mu_p_by_p0}
\end{figure}

The steady-state pressure $P$ at $t = 500\text{s}$ is normalized by the initial bulk pressure $P_0$ at $t = 0$, giving the scaled pressure $\langle P / P_0 \rangle_z$. \autoref{fig:AR_mu_p_by_p0} shows $\langle P/P_{0} \rangle_z$ and the packing density $\langle \phi \rangle_z$ as functions of $y$, both averaged over the middle $z$-layer in the steady state. The grey shaded region indicates the shear band width, averaged over the $z$-direction.

At small $\mu$ ($\approx 0.01$), $\langle P / P_0 \rangle_z$ is nearly uniform across $y$. As $\mu$ increases, the pressure within the shear band increases relative to its surroundings. This can be understood from the width of the shear band with increasing friction. 

The shear band is wide at low $\mu$ \cite{singh2024shear}, allowing particles to align and compact over a larger area. This leads to a high packing density inside the shear band, as seen in \autoref{fig:AR_mu_p_by_p0}(b). Because the deformation is spread over a wide shear zone, the stress does not localize, and the pressure remains nearly uniform across $y$. With increasing $\mu$, the shear band narrows, restricting particle rearrangement and decreasing the packing density within the shear band (\autoref{fig:AR_mu_p_by_p0}(b)). At the same time, the local pressure increases. This results from stress localization within the narrow shear band. The same forces are transmitted through fewer contacts over a smaller region, resulting in a large local normal stress. \autoref{fig:AR_mu_p_by_p0}(b) shows only the shear-band region and adjacent bulk, where dilation is strongest, rather than the full $y$-range.

Conversely, spherical particles maintain a nearly uniform pressure across all $\mu$ values as shown in \autoref{fig:AR_mu_p_by_p0}(c). Because of their shape symmetry, spherical particles rearrange easily under shear, preventing stress localization. As a result, the pressure remains nearly uniform across $y$, for all $\mu$ values. This indicates that both particle shape and friction affect the pressure distribution in sheared granular materials.

\section{Conclusions}
We investigated how particle shape ($\text{AR}$) and interparticle friction ($\mu$) influence the pressure field in granular materials sheared in a linear split-bottom shear cell. Coarse-grained pressure and stress fields were used to compare simulation results for systems composed of spherical and elongated particles under shear.

Elongated particles develop pressure anisotropy due to shear-induced alignment along the flow direction. This alignment decreases interparticle spacing and increases local packing density, fostering both lateral and vertical normal stress components, resulting in higher pressure. This pressure anisotropy builds up gradually as shear progresses and saturates in the steady state.

Large interparticle friction accentuates this effect in elongated particles by narrowing the shear band and concentrating stress over a smaller region. The same forces act through fewer contacts over a smaller area, leading to higher local stress and pressure. 

In contrast, spherical particles do not align and are weakly influenced by friction. Their symmetric shape allows them to rotate, translate, and rearrange easily, maintaining a uniform pressure distribution before and after shear. 

These findings indicate a relationship between particle shape, microstructure, and stress distribution. The behavior of elongated particles under shear shows a resemblance to the Weissenberg effect observed in non-Newtonian fluids, where normal stress differences induce a radial pressure gradient. Likewise, elongated particles develop a varied surface profile and a pressure difference between the shear band and the surrounding region due to alignment and stress anisotropy. In contrast, spherical particles behave more like Newtonian fluids, showing uniform pressure without stress localization.

\section{Acknowledgement}
We thank Dr. Patric Müller for stimulating discussion. Financial support through the DFG grant for the Granular Weissenberg Effect (Project No. PO472/40-1) is acknowledged.



\bibliography{references}

@article{rahim2025impact,
  title={Impact of friction and grain shape on the morphology of sheared granular media},
  author={Rahim, Huzaif and Roy, Sudeshna and P{\"o}schel, Thorsten},
  journal={arXiv preprint arXiv:2502.13739},
  year={2025},
 doi = {10.48550/arXiv.2502.13739}
}

@inproceedings{thorens2021taming,
  title={Taming the Janssen effect},
  author={Thorens, Louison and M{\aa}l{\o}y, Knut J{\o}rgen and Bourgoin, Micka{\"e}l and Santucci, St{\'e}phane},
  booktitle={EPJ Web of Conferences},
  volume={249},
  pages={08004},
  year={2021},
  organization={EDP Sciences},
 doi = {10.1051/epjconf/202124908004}
}

@article{krishnaraj2016dilation,
  title={A dilation-driven vortex flow in sheared granular materials explains a rheometric anomaly},
  author={Krishnaraj, KP and Nott, Prabhu R},
  journal={Nature Communications},
  volume={7},
  number={1},
  pages={10630},
  year={2016},
  publisher={Nature Publishing Group UK London},
  doi = {10.1038/ncomms10630}
}

@article{strobl2016exact,
  title={Exact calculation of the overlap volume of spheres and mesh elements},
  author={Strobl, Severin and Formella, Arno and P{\"o}schel, Thorsten},
  journal={Journal of Computational Physics},
  volume={311},
  pages={158--172},
  year={2016},
  publisher={Elsevier},
  doi = {10.1016/j.jcp.2016.02.003}
}

@article{pol2022kinematics,
  title={Kinematics and shear-induced alignment in confined granular flows of elongated particles},
  author={Pol, Antonio and Artoni, Riccardo and Richard, Patrick and da Concei{\c{c}}{\~a}o, Paulo Ricardo Nunes and Gabrieli, Fabio},
  journal={New Journal of Physics},
  volume={24},
  number={7},
  pages={073018},
  year={2022},
  publisher={IOP Publishing},
 doi = {10.1088/1367-2630/ac7d6d}
}

@article{borzsonyi2012orientational,
  title={Orientational order and alignment of elongated particles induced by shear},
  author={B{\"o}rzs{\"o}nyi, Tam{\'a}s and Szab{\'o}, Bal{\'a}zs and T{\"o}r{\"o}s, G{\'a}bor and Wegner, Sandra and T{\"o}r{\"o}k, J{\'a}nos and Somfai, Ell{\'a}k and Bien, Tomasz and Stannarius, Ralf},
  journal={Physical Review Letters},
  volume={108},
  number={22},
  pages={228302},
  year={2012},
  publisher={APS},
 doi = {10.1103/PhysRevLett.108.228302}
}

@article{fu2011fabric,
  title={Fabric evolution within shear bands of granular materials and its relation to critical state theory},
  author={Fu, Pengcheng and Dafalias, Yannis F},
  journal={International Journal for numerical and analytical methods in geomechanics},
  volume={35},
  number={18},
  pages={1918--1948},
  year={2011},
  publisher={Wiley Online Library},
 doi = {10.1002/nag.988}
}

@article{liu2018application,
  title={Application of the two-fluid model with kinetic theory of granular flow in liquid--solid fluidized beds},
  author={Liu, Guodong},
  journal={Granularity in Materials Science},
  volume={2},
  year={2018},
  publisher={InTech Rijeka},
 doi = {10.5772/intechopen.79696}
}

@article{vescovi2016merging,
  title={Merging fluid and solid granular behavior},
  author={Vescovi, Dalila and Luding, Stefan},
  journal={Soft matter},
  volume={12},
  number={41},
  pages={8616--8628},
  year={2016},
  publisher={Royal Society of Chemistry},
 doi = {10.1039/c6sm01444e}
}

@article{lei2019side,
  title={Side-wall pressure distribution of cylindrical granular containers with flat bottom},
  author={Lei, Yusheng and Ma, Qinwei and Shi, Qingfan},
  journal={Powder Technology},
  volume={353},
  pages={57--63},
  year={2019},
  publisher={Elsevier},
 doi = {10.1016/j.powtec.2019.05.003}
}

@article{liu2024effect,
  title={Effect of Particle Shape on Contact Network and Shear-Induced Anisotropy of Granular Assemblies: A DEM Perspective},
  author={Liu, Yang and Liu, Xu and Ren, Jinlan},
  journal={Journal of Geotechnical and Geoenvironmental Engineering},
  volume={150},
  number={3},
  pages={04023142},
  year={2024},
  publisher={American Society of Civil Engineers},
 doi = {10.1061/JGGEFK.GTENG-11762}
}

@article{kanzaki2011stress,
  title={Stress distribution of faceted particles in a silo after its partial discharge},
  author={Kanzaki, T and Acevedo, M and Zuriguel, Iker and Pagonabarraga, Ignacio and Maza, Diego and Hidalgo, RC},
  journal={The European Physical Journal E},
  volume={34},
  pages={1--8},
  year={2011},
  publisher={Springer},
 doi = {10.1140/epje/i2011-11133-5}
}

@article{hidalgo2009role,
  title={Role of particle shape on the stress propagation in granular packings},
  author={Hidalgo, Ra{\'u}l Cruz and Zuriguel, Iker and Maza, Diego and Pagonabarraga, Ignacio},
  journal={Physical review letters},
  volume={103},
  number={11},
  pages={118001},
  year={2009},
  publisher={APS},
 doi = {10.1103/PhysRevLett.103.118001}
}

@article{zhu2023discrete,
  title={Discrete Element Simple Shear Test Considering Particle Shape},
  author={Zhu, Houying and Li, Xuefeng and Lv, Longlong and Yuan, Qi},
  journal={Applied Sciences},
  volume={13},
  number={20},
  pages={11382},
  year={2023},
  publisher={MDPI},
 doi = {10.3390/app132011382}
}

@article{sperl2006experiments,
  title={Experiments on corn pressure in silo cells--translation and comment of Janssen's paper from 1895},
  author={Sperl, Matthias},
  journal={Granular Matter},
  volume={8},
  number={2},
  pages={59--65},
  year={2006},
  publisher={Springer},
 doi = {10.1007/s10035-005-0224-z}
}

@article{singh2024shear,
  title={Shear zones in granular mixtures of hard and soft particles with high and low friction},
  author={Singh, Aditya Pratap and Angelidakis, Vasileios and P{\"o}schel, Thorsten and Roy, Sudeshna},
  journal={Soft Matter},
  volume={20},
  number={14},
  pages={3118--3130},
  year={2024},
  publisher={Royal Society of Chemistry},
 doi = {10.1039/D4SM00100A}
}

@article{campbell2011elastic,
  title={Elastic granular flows of ellipsoidal particles},
  author={Campbell, Charles S},
  journal={Physics of Fluids},
  volume={23},
  number={1},
  pages={013306},
  year={2011},
  publisher={AIP Publishing},
 doi = {10.1063/1.3546037}
}

@article{fischer2016heaping,
  title={Heaping and secondary flows in sheared granular materials},
  author={Fischer, David and B{\"o}rzs{\"o}nyi, Tam{\'a}s and Nasato, Daniel S and P{\"o}schel, Thorsten and Stannarius, Ralf},
  journal={New Journal of Physics},
  volume={18},
  number={11},
  pages={113006},
  year={2016},
  publisher={IOP Publishing},
 doi = {10.1088/1367-2630/18/11/113006}
}

@article{wortel2015heaping,
  title={Heaping, secondary flows and broken symmetry in flows of elongated granular particles},
  author={Wortel, Geert and B{\"o}rzs{\"o}nyi, Tam{\'a}s and Somfai, Ell{\'a}k and Wegner, Sandra and Szab{\'o}, Bal{\'a}zs and Stannarius, Ralf and van Hecke, Martin},
  journal={Soft Matter},
  volume={11},
  number={13},
  pages={2570--2576},
  year={2015},
  publisher={Royal Society of Chemistry},
 doi = {10.1039/C4SM02534B}
}

@article{dijksman2010granular,
  title={Granular flows in split-bottom geometries},
  author={Dijksman, Joshua A and Van Hecke, Martin},
  journal={Soft Matter},
  volume={6},
  number={13},
  pages={2901--2907},
  year={2010},
  publisher={Royal Society of Chemistry},
 doi = {10.1039/B925110C}
}

@article{wegner2014effects,
  title={Effects of grain shape on packing and dilatancy of sheared granular materials},
  author={Wegner, Sandra and Stannarius, Ralf and Boese, Axel and Rose, Georg and Szabo, Balazs and Somfai, Ellak and B{\"o}rzs{\"o}nyi, Tam{\'a}s},
  journal={Soft Matter},
  volume={10},
  number={28},
  pages={5157--5167},
  year={2014},
  publisher={Royal Society of Chemistry},
 doi = {10.1039/C4SM00838C}
}

@article{Buchholtz:1993,
    author = {P\"oschel, T. and Buchholtz, V.},
    title = {Static friction phenomena in granular materials: {C}oulomb law versus particle geometry},
    journal = {Physical Review Letters},
    volume = {71},
    year ={1993},
    pages = {3963-3966},
    doi = {10.1103/PhysRevLett.71.3963}
}

@article{roy2021drift,
  title={Drift-diffusive liquid migration in partly saturated sheared granular media},
  author={Roy, S and Luding, S and Den Otter, WK and Thornton, AR and Weinhart, T},
  journal={Journal of Fluid Mechanics},
  volume={915},
  pages={A30},
  year={2021},
  publisher={Cambridge University Press},
 doi = {10.1017/jfm.2021.30}
}

@article{ries2007shear,
  title={Shear zones in granular media: three-dimensional contact dynamics simulation},
  author={Ries, Alexander and Wolf, Dietrich E and Unger, Tam{\'a}s},
  journal={Physical Review E},
  volume={76},
  number={5},
  pages={051301},
  year={2007},
  publisher={APS},
 doi = {10.1103/PhysRevE.76.051301}
}

@article{abou2004three,
  title={Three-dimensional particle shape descriptors for computer simulation of non-spherical particulate assemblies},
  author={Abou-Chakra, Hd and Baxter, Jd and T{\"u}z{\"u}n, Ud},
  journal={Advanced Powder Technology},
  volume={15},
  number={1},
  pages={63--77},
  year={2004},
  publisher={Elsevier},
 doi = {10.1163/15685520460740070}
}

@article{kodam2009force,
  title={Force model considerations for glued-sphere discrete element method simulations},
  author={Kodam, Madhusudhan and Bharadwaj, Rahul and Curtis, Jennifer and Hancock, Bruno and Wassgren, Carl},
  journal={Chemical Engineering Science},
  volume={64},
  number={15},
  pages={3466--3475},
  year={2009},
  publisher={Elsevier},
 doi = {10.1016/j.ces.2009.04.025}
}

@article{cabiscol2018calibration,
  title={Calibration and interpretation of DEM parameters for simulations of cylindrical tablets with multi-sphere approach},
  author={Cabiscol, Ramon and Finke, Jan Henrik and Kwade, Arno},
  journal={Powder Technology},
  volume={327},
  pages={232--245},
  year={2018},
  publisher={Elsevier},
 doi = {10.1016/j.powtec.2017.12.041}
}

@article{muller2011collision,
  title={Collision of viscoelastic spheres: Compact expressions for the coefficient of normal restitution},
  author={M{\"u}ller, Patric and P{\"o}schel, Thorsten},
  journal={Physical Review E},
  volume={84},
  number={2},
  pages={021302},
  year={2011},
  publisher={APS},
 doi = {10.1103/PhysRevE.84.021302}
}

@article{parteli2016particle,
  title={Particle-based simulation of powder application in additive manufacturing},
  author={Parteli, Eric JR and P{\"o}schel, Thorsten},
  journal={Powder Technology},
  volume={288},
  pages={96--102},
  year={2016},
  publisher={Elsevier},
 doi = {10.1016/j.powtec.2015.10.035}
}

@article{nagy2023flow,
  title={Flow of asymmetric elongated particles},
  author={Nagy, Viktor and Fan, Bo and Somfai, Ell{\'a}k and Stannarius, Ralf and B{\"o}rzs{\"o}nyi, Tam{\'a}s},
  journal={Journal of Statistical Mechanics: Theory and Experiment},
  volume={2023},
  number={11},
  pages={113201},
  year={2023},
  publisher={IOP Publishing},
 doi = {10.1088/1742-5468/ad0831}
}

@article{mindlin1949compliance,
  title={Compliance of elastic bodies in contact},
  author={Mindlin, Raymond David},
  journal={Journal of Applied Mechanics},
  volume={16},
  pages={259--268},
  year={1949},
  publisher={American Society of Mechanical Engineers},
  doi={10.1115/1.4009973}
}

@article{weinhart2016influence,
  title={Influence of coarse-graining parameters on the analysis of DEM simulations of silo flow},
  author={Weinhart, Thomas and Labra, Carlos and Luding, Stefan and Ooi, Jin Y},
  journal={Powder technology},
  volume={293},
  pages={138--148},
  year={2016},
  publisher={Elsevier},
 doi = {10.1016/j.powtec.2015.11.052}
}

@article{singh2015role,
  title={The role of gravity or pressure and contact stiffness in granular rheology},
  author={Singh, Abhinendra and Magnanimo, Vanessa and Saitoh, Kuniyasu and Luding, Stefan},
  journal={New Journal of Physics},
  volume={17},
  number={4},
  pages={043028},
  year={2015},
  publisher={IOP Publishing},
 doi = {10.1088/1367-2630/17/4/043028}
}

@article{depken2006continuum,
  title={Continuum approach to wide shear zones in quasistatic granular matter},
  author={Depken, Martin and van Saarloos, Wim and van Hecke, Martin},
  journal={Physical Review E},
  volume={73},
  number={3},
  pages={031302},
  year={2006},
  publisher={APS},
 doi = {10.1103/PhysRevE.73.031302}
}

@article{hosseinpoor2021rheo,
  title={Rheo-morphological investigation of Reynolds dilatancy and its effect on pumpability of self-consolidating concrete},
  author={Hosseinpoor, Masoud and Koura, Baba-Issa Ouro and Yahia, Ammar},
  journal={Cement and Concrete Composites},
  volume={117},
  pages={103912},
  year={2021},
  publisher={Elsevier},
 doi = {10.1016/j.cemconcomp.2020.103912}
}

@article{fenistein2003wide,
  title={Wide shear zones in granular bulk flow},
  author={Fenistein, Denis and Van Hecke, Martin},
  journal={Nature},
  volume={425},
  number={6955},
  pages={256--256},
  year={2003},
  publisher={Nature Publishing Group UK London},
 doi = {10.1038/425256a}
}

@article{dsouza2021dilatancy,
  title={Dilatancy-driven secondary flows in dense granular materials},
  author={Dsouza, Peter Varun and Nott, Prabhu R},
  journal={Journal of Fluid Mechanics},
  volume={914},
  pages={A36},
  year={2021},
  publisher={Cambridge University Press},
 doi = {10.1017/jfm.2020.1029}
}

@inproceedings{dsouza2017secondary,
  title={Secondary flows in slow granular flows},
  author={Dsouza, Peter Varun and Krishnaraj, KP and Nott, Prabhu R},
  booktitle={EPJ Web of Conferences},
  volume={140},
  pages={03028},
  year={2017},
  organization={EDP Sciences},
 doi = {10.1051/epjconf/201714003028}
}

@article{jaeger1996granular,
  title={Granular solids, liquids, and gases},
  author={Jaeger, Heinrich M and Nagel, Sidney R and Behringer, Robert P},
  journal={Reviews of Modern Physics},
  volume={68},
  number={4},
  pages={1259},
  year={1996},
  publisher={APS},
  doi = {10.1103/RevModPhys.68.1259}
}

@article{anthony2005influence,
  title={Influence of particle characteristics on granular friction},
  author={Anthony, Jennifer L and Marone, Chris},
  journal={Journal of Geophysical Research: Solid Earth},
  volume={110},
  number={B8},
  pages={B08409},
  year={2005},
  publisher={Wiley Online Library},
 doi = {10.1029/2004JB003399}
}

@article{cheng2000dynamic,
  title={Dynamic simulation of random packing of spherical particles},
  author={Cheng, YF and Guo, SJ and Lai, HY},
  journal={Powder Technology},
  volume={107},
  number={1-2},
  pages={123--130},
  year={2000},
  publisher={Elsevier},
 doi = {10.1016/S0032-5910(99)00178-3}
}

@article{yu1998prediction,
  title={Prediction of the porosity of particle mixtures},
  author={Yu, AB and Zou, RP},
  journal={KONA Powder and Particle Journal},
  volume={16},
  number={0},
  pages={68--81},
  year={1998},
  publisher={Hosokawa Powder Technology Foundation},
 doi = {10.14356/kona.1998010}
}

@article{reynolds1885lvii,
  title={LVII. On the dilatancy of media composed of rigid particles in contact. With experimental illustrations},
  author={Reynolds, Osborne},
  journal={The London, Edinburgh, and Dublin Philosophical Magazine and Journal of Science},
  volume={20},
  number={127},
  pages={469--481},
  year={1885},
  publisher={Taylor \& Francis},
 doi = {10.1080/14786448508627791}
}

@article{rahim2024alignment,
  title={Alignment-induced depression and shear thinning in granular matter of nonspherical particles},
  author={Rahim, Huzaif and Angelidakis, Vasileios and P{\"o}schel, Thorsten and Roy, Sudeshna},
  journal={Physical Review Fluids},
  volume={9},
  number={11},
  pages={114304},
  year={2024},
  publisher={APS},
 doi = {https://doi.org/10.1103/PhysRevFluids.9.114304}
}

@article{luding2008introduction,
  title={Introduction to discrete element methods: basic of contact force models and how to perform the micro-macro transition to continuum theory},
  author={Luding, Stefan},
  journal={European Journal of Environmental and Civil Engineering},
  volume={12},
  number={7-8},
  pages={785--826},
  year={2008},
  publisher={Taylor \& Francis},
 doi = {10.3166/ejece.12.785-826}
}

@article{thornton2000numerical,
  title={Numerical simulations of deviatoric shear deformation of granular media},
  author={Thornton, Colin},
  journal={Geotechnique},
  volume={50},
  number={1},
  pages={43--53},
  year={2000},
  publisher={Thomas Telford Ltd},
  doi = {10.1680/geot.2000.50.1.43}
}

@article{cundall1979discrete,
  title={A discrete numerical model for granular assemblies},
  author={Cundall, Peter A and Strack, Otto DL},
  journal={Geotechnique},
  volume={29},
  number={1},
  pages={47--65},
  year={1979},
  publisher={Thomas Telford Ltd},
  doi = {10.1680/geot.1979.29.1.47}
}

@article{BSHP:1996,
  title = {Model for collisions in granular gases},
  author = {Brilliantov, Nikolai V. and Spahn, Frank and Hertzsch, Jan-Martin and P\"oschel, Thorsten},
  journal = {Physical Review E},
  volume = {53},
  pages = {5382--5392},
  year = {1996},
  doi = {10.1103/PhysRevE.53.5382}
}

@article{aguilar2025janssen,
  title={Janssen effect in submerged granular columns},
  author={Aguilar-Gonz{\'a}lez, M and Maza, D and Pacheco-V{\'a}zquez, F},
  journal={Soft Matter},
  volume={21},
  number={31},
  pages={6234--6242},
  year={2025},
  publisher={Royal Society of Chemistry},
  doi = {10.1039/D5SM00523J}
}

\end{document}